\documentclass[fleqn,usenatbib]{mnras}
\setcitestyle{notesep={ }} 

\usepackage{newtxtext,newtxmath}

\usepackage[T1]{fontenc}

\usepackage{longtable}

\DeclareRobustCommand{\VAN}[3]{#2}
\let\VANthebibliography\thebibliography
\def\thebibliography{\DeclareRobustCommand{\VAN}[3]{##3}\VANthebibliography}

\usepackage{graphicx}	
\usepackage{amsmath}	
\usepackage{booktabs}   
\usepackage{xcolor}       

\newcommand{\dotdeg}{\rlap{.}^\circ}
  {\color{red}}%
  {}

\title[Testing the Cosmological Principle]{Testing the Cosmological Principle:  
On the Time Dilation \\ of Distant Sources}

\author[Oayda \& Lewis]{Oliver~T.~Oayda\thanks{E-mail: ooay3125@uni.sydney.edu.au (OTO)} \& Geraint~F.~Lewis\thanks{E-mail: geraint.lewis@sydney.edu.au (GFL)}
\\
Sydney Institute for Astronomy, School of Physics, A28, The University of Sydney, NSW 2006, Australia\\
}

\date{Accepted 2023 May 10. Received 2023 March 20; in original form 2023 January 15}

\pubyear{2023}

\begin{document}
\label{firstpage}
\pagerange{\pageref{firstpage}--\pageref{lastpage}}
\maketitle

\begin{abstract}
We present a novel test of the cosmological principle: the idea that, 
on sufficiently large scales, the universe should 
appear homogeneous and isotropic to observers comoving with the Hubble flow. 
This is a fundamental assumption in modern cosmology,
underpinning the use of the
Friedmann-Lema\^{i}tre-Robertson-Walker metric as part
of the concordance $\Lambda$CDM paradigm.
However, the observed dipole imprinted on the Cosmic Microwave 
Background (CMB) is interpreted as our departure from the Hubble flow, 
and such a proper motion will induce a directionally-dependent 
time dilation over the sky.
We illustrate the feasibility of detection of this `time dilation dipole'
and sketch the practical steps involved in its extraction from a catalogue
of sources with intrinsic time-scales.
In essence, whilst the scale of this dilation is small, being of order of $0.1\%$, it will
in principle be detectable in large scale surveys of variable cosmological sources, 
such as quasars and supernovae.
The degree of alignment of the time dilation dipole with the kinematic 
dipole derived from the CMB will provide a new 
assessment of the cosmological principle, and address the tension in 
dipole measures from other observations.
\end{abstract}




\section{Introduction}
\label{Sec:introduction}
The cosmological principle, which posits that large-scale views of the universe should appear homogeneous 
and isotropic to comoving observers, underpins our modern cosmological ideas 
\citep[see][]{2000csu..book.....H}.
Namely, the key proposition of homogeneity and isotropy is incorporated within the Friedmann-Lema\^{i}tre-Robertson-Walker (FLRW) metric of spacetime, as employed by the current concordance cosmological model, $\Lambda$CDM.
As guided by the cosmological principle, comoving observers define a cosmological rest frame, which is conventionally taken to correspond to the frame in which the Cosmic Microwave Background (CMB) is perceived as maximally isotropic with temperature anisotropies $\Delta T / T \approx 10^{-5}$ \citep{maartens2011}.
The fact that the CMB is observed to exhibit a dipole has been interpreted as our departure 
from the local Hubble flow -- that is, Earth's deviation from the cosmic rest frame -- with a peculiar velocity of $369.82\pm0.11\,\text{km}\,\text{s}^{-1}$ 
towards $(l,b) = (264\dotdeg0.21, 48\dotdeg253)$ \citep{2020AA...641A...1P}.
If this interpretation is correct, then this dipole should be imprinted on other observations
of cosmological sources \citep[e.g.][]{1984MNRAS.206..377E}, although more recently there
has been a claimed tension between the magnitude and direction of cosmic dipoles when compared
to the CMB \citep[see][for a recent review]{2022arXiv220705765A}. Insofar that this evinces disagreement between the standard of rest defined by the CMB and other sources of cosmological origin, some have proposed that the tension represents a challenge to the cosmological principle and hence the foundations 
of modern relativistic cosmologies.

Motivated by this tension, we propose a novel test of the cosmological principle, formulated through the observed time dilation of distant sources.
In Section~\ref{Sec:cosmological}, we briefly
outline recent determinations of cosmic dipoles from various cosmological observations, 
indicating the degree of the current tension.
In Section~\ref{Sec:timedilation}, we outline
our approach, demonstrated through the numerical simulation of a mock catalogue of sources each carrying an intrinsic timescale.
The effects of kinematic time dilation are then imprinted onto the sources, and, following the procedure of Bayesian hypothesis testing, Earth's peculiar motion is recovered.
This serves to elucidate the key steps involved in our probe of the cosmological principle if it were to be performed in reality.
In Section~\ref{Sec:observations} we discuss the observational 
prospects given future surveys.
We summarise this work and present our conclusions in Section~\ref{Sec:conclusions}.

\section{Testing the Cosmological Principle}
\label{Sec:cosmological}
\subsection{Matter dipole studies}
A key way of assessing the consistency of the cosmological principle -- which was initially postulated \textit{a priori} by \citet{milne1935} -- is by determining the degree of concordance between the CMB dipole and the `matter dipole'.
In particular, we can apply a local Lorentz boost to transform into a frame in which the CMB exhibits no dipole.
This is postulated as the cosmic rest frame.
If this interpretation -- typically called the `kinematic interpretation' of the CMB dipole -- holds, then distributions of matter should be perceived as isotropic to an observer (the fundamental observer) in this frame.
\citet{1984MNRAS.206..377E} proposed a test along these lines through observations of radio galaxy counts.
In the cosmic rest frame, the distribution of radio galaxies is assumed to be uniform, as per the cosmological principle.
However, Earth's peculiar velocity $v \ll c$ induces a directionally-dependent Doppler shift, leading to a dipole anisotropy in source number count per unit solid angle.
If the flux density of the sources are described by a power law $S \propto v^{-\alpha}$ and an integral source count per unit solid angle above a limiting flux density $S_\nu$ given by $d N / d \Omega \propto S_\nu^{-x}$, then the authors expect a dipole of amplitude
\begin{equation}
    \mathcal{D} = [2 + x(1 + \alpha)] \beta
\end{equation}
where $\beta = v / c$. Catalogues of radio galaxies, collated after the formulation of this test, offer a means to determine the direction and magnitude of the matter dipole and check its degree of concordance with the CMB dipole.
Despite earlier indications of consistency with the cosmological principle as in \cite{blake2002}, more recent studies across different groups generally find a matter dipole which aligns with the CMB dipole, yet has an amplitude greater than expected \citep[see e.g.][]{colin2017,bengaly2018,singal2019}.
A more recent study performed in \citet{siewert2021} corroborates this proposition, and Appendix A therein provides a summary of the previous key results in the literature.
We also defer the reader to \citet{2022arXiv220705765A} for a full review of these studies in a more general setting amongst other probes of the cosmological principle, not necessarily related to distributions of matter arising from an Earth-bound observer's peculiar velocity.
We note, however, that this measured inconsistency contrasts with the recent work of \citet{darling2022}.
There, the author interrogated the distribution and brightness of radio galaxies with more contemporary catalogues, namely the Very Large Array Sky Survey \citep[VLASS;][]{Lacy2020} and the Rapid Australian Square Kilometre Array Pathfinder \citep[RACS;][]{mcconnell2020}, finding these metrics to be substantially consistent with motion aligned with the CMB dipole and of the same magnitude.
Yet it is possible, as the author considered, that their finding is not necessarily inconsistent with the excessive dipole measurements from other works, for instance in \citet{2021ApJ...908L..51S} which we turn to below.
This is because of the high-velocity tail of their posterior probability distribution for the dipole amplitude, which introduces significant uncertainty to the measurement.

Of recent interest, and arguably representing one of the most significant challenges to the cosmological principle, are the findings of \citet{2021ApJ...908L..51S}.
There, the authors carried out the test of \citet{1984MNRAS.206..377E} with a catalogue of $\approx$ 1 million quasars observed by the \textit{Wide-field Infrared Survey Explorer} \citep[WISE;][]{wright2020}, as forming part of the CatWISE2020 catalgoue \citep{marocco2021}.
Using a least-squares estimator, the authors reported a number count dipole of magnitude $\mathcal{D} = 0.01554$ in the direction $(l,b)=(238\dotdeg2,28\dotdeg8)$.
This is twice as large as expected and $27\dotdeg8$ off-alignment in light of the direction and magnitude of the CMB dipole respectively.
\citet{secrest2022} corroborated these findings in a joint analysis of the CatWISE2020 quasars and radio galaxies from the NRAO VLA Sky Survey \citep[NVSS;][]{nvss-survey}, reporting a $5.2 \sigma$ joint significance with which the null hypothesis of concordance between the CMB and matter dipoles was rejected.
More recently, \citet{kothari2022} independently verified the excessive dipole from the WISE quasars, although reported that the mean spectral index and flux density of the sources were consistent with isotropy.
Further, \citet{dam2022} revisited the CatWISE2020 quasars, in this case employing the methodology of Bayesian statistics to check the number count dipole.
The authors reached the conclusion that, independent of the choice of statistical language, the magnitude of the dipole is in excess, while the direction is consistent with the CMB dipole.
This is still the case even after considering various systematics which could contaminate the dipole measurement, like the impact of the galactic plane mask used originally in \cite{2021ApJ...908L..51S}.
Thus the CatWISE2020 quasar sample remains an outstanding anomaly in light of the prevailing $\Lambda$CDM paradigm.

In a similar vein, probes of the matter dipole have been carried out with SNe Ia, in which one postulates under the cosmological principle that the supernovae form a cosmic rest frame.
Earth's peculiar motion, however, will shift observed supernovae in a magnitude-redshift plot depending on the angle between the source and the direction of motion of Earth; sources in the forward and backward hemispheres will be systematically offset in such a way to allow determination of Earth's peculiar velocity.
Employing this method, \citet{singal2022} reported that the supernovae dipole was generally in agreement with the CMB dipole at a $\lesssim 2 \sigma$ level, but is $\approx$ 4 times larger in magnitude than the CMB dipole, which suggests some degree of synergy with the aforementioned radio galaxy and quasar tests.
However, this is inconsistent with the findings of \citet{horstmann2022}, in which the authors reported a dipole magnitude less than that of the CMB.

\subsection{Curvature studies}
In the foregoing tests, Earth's peculiar motion is interpreted as a local perturbation to the homogeneous and isotropic FLRW metric.
In this manner, the probes assess whether or not the inhomogeneities are consistent with the FLRW paradigm, and are generally independent of a cosmological model apart from the initial assumption of the cosmological principle \citep{2022arXiv220705765A}.
However, we note that there are additional model-independent ways to probe the FLRW metric, for example by measuring the constancy of the spatial curvature parameter $\Omega_k$ across different measurements.
Recent work has considered using pairs of strongly-lensed gravitational waves and their corresponding electromagnetic signal to find comoving distances between the observer, source and lens \citep[see e.g.][]{cao2019,zhou2020,liu2020}.
Employing the distance-sum rule for null geodesics in the FLRW metric then enables the determination of $\Omega_k$ for source-lens systems at different redshifts.
If they are not constant for different redshift pairs, this suggests against the accuracy of the FLRW paradigm \citep{rasanen2015}.
At this stage, it is too early to say whether or not there is any evolution in spatial curvature, but `considerable advances' are expected in precision gravitational wave astronomy in the coming decade \citep{2022arXiv220705765A}. Thus, this type of curvature test could soon be a feasible way to probe the cosmological principle.

\subsection{Current tension}
\noindent What we have discussed above is not an exhaustive survey of the literature.
However, the differing and in some cases conflicting results illustrate the emerging discord between the CMB dipole and other cosmic dipoles.
Arguably, at least in the context of the radio galaxy/quasar studies, the evidence suggests that the matter dipole is larger in magnitude than anticipated.
Notwithstanding the possibility of systematics which underlie the various catalogues of sources used, this presents a serious challenge to the cosmological principle and its relevance in the prevailing assumptions of modern cosmology.
Namely, anomalous results have the role of gradually eroding the $\Lambda$CDM framework; as the authors of \citet{2022arXiv220705765A} note (their emphasis), `[I]t is extremely difficult to confirm FLRW, since \textit{this is the default that one recovers either with poor quality data or insensitive cosmological probes}.
For this reason, one can only hope to falsify FLRW by focusing on the anomalies.'
The key point is that substantial evidential weight from a multitude of independent tests is necessary to either confirm or reject the assumption of homogeneity and isotropy.
In this manner, the formulation of new, independent tests is critical for adding to the nascent body of evidence used to assess the accuracy of the cosmological principle, especially when there is already conflict across different studies.
This has motivated us to devise a new test of Earth's peculiar motion determined from variable astronomical sources.
This test functions similarly to the matter dipole tests mentioned earlier, but critically focuses on the intrinsic timescale associated with the variability of each source.
Importantly, these timescales will be time dilated as predicted by special relativity, enabling a comparison between the direction and magnitude of Earth's inferred motion and the CMB dipole, which we describe in Section~\ref{sub:princ-kin-dil}.
Hence, we present an additional means to probe the cosmological principle.
We envision this test being used to better understand the tension in the matter dipole studies, adding an additional check on the conflicting results.

\section{Approach}
\label{Sec:timedilation}
We first describe time dilation, contrasting the cosmological case from the kinematic one so as to clarify the physical principles which underpin our novel probe. We then explain in more detail the necessary steps of our test.

\subsection{Time dilation}
\label{sub:time-dilation}

\subsubsection{Cosmological time dilation}
\label{sub:cosm-time-dil}
It is well known that the expansion of the universe induces a cosmological 
time dilation, which is equal to $(1+z)$ and has already been observed in the characteristic
time-scale of cosmologically distant supernovae \citep{1997ASIC..486..777G,2008ApJ...682..724B}, 
as well as hinted at in gamma ray bursts \citep[GRBs;][]{1997ApJ...483L..25C,2013ApJ...778L..11Z,2022JCAP...02..010S}.
The situation is not identical with the other prominently varying cosmological sources, namely quasars, with 
recent claims that they do not display the expected time dilation as a function of their 
redshifts \citep[see for example][for a recent contribution]{2022MNRAS.512.5706H}. However, whilst the samples of monitored quasars is substantial, namely being of order one thousand objects, 
and the monitoring period is more than 25 years, the cadence of sampling is generally poor, 
with only 3-4 observations per year. 
More intensive observations of quasars have revealed multi-scale variability that is 
characterised by a damped random walk \citep[DRW; see e.g.][]{kelly2009,2013ApJ...765..106Z,2014IAUS..304..395I}
and it may simply be that existing large samples of monitored quasars are insufficiently 
sampled to accurately determine the characteristic properties of the variability. We will
return to this point in Section~\ref{Sec:observations}.

\begin{figure*}
    \centering
    \includegraphics[height=6cm]{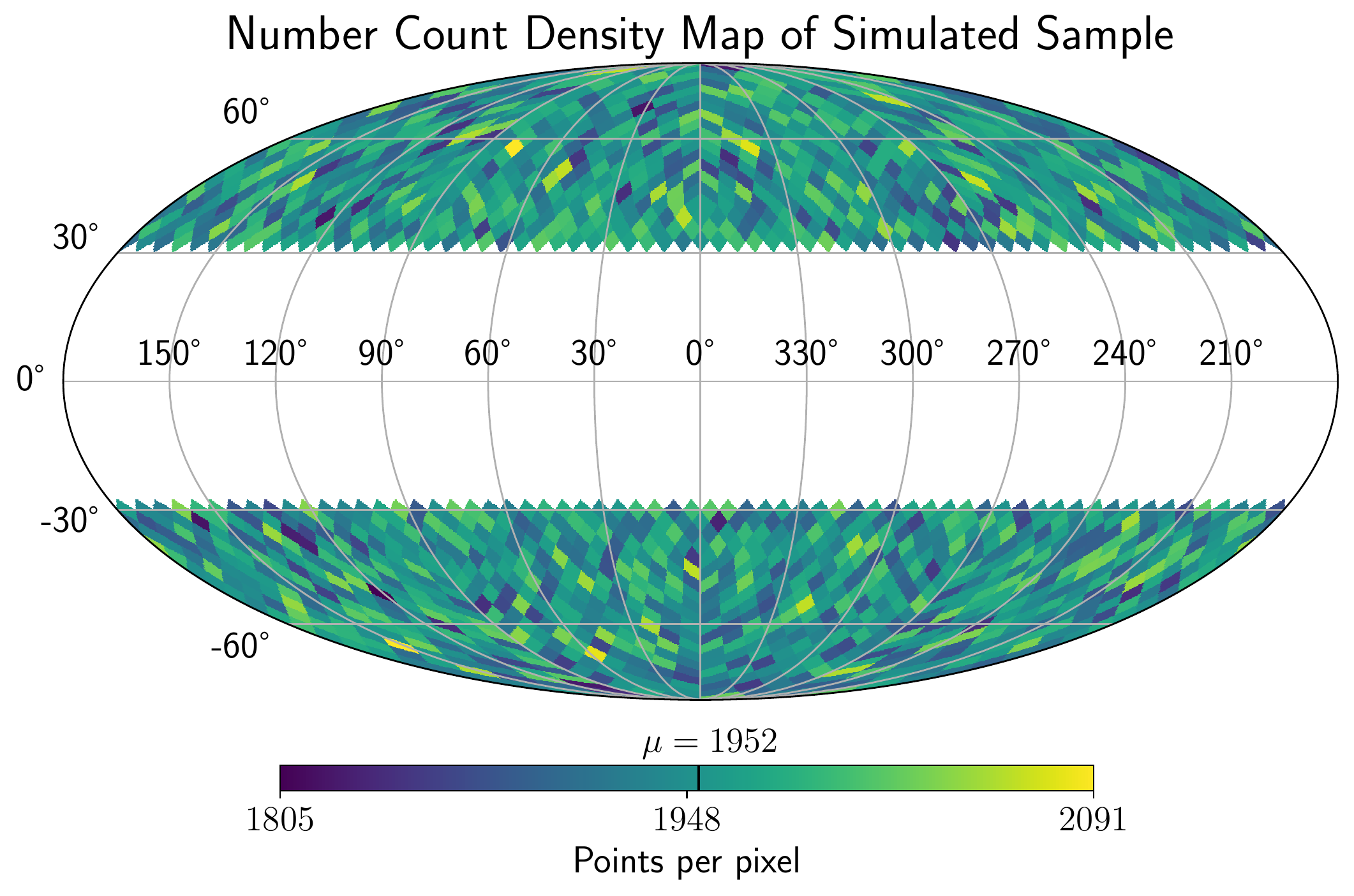}
    \hfill
    \includegraphics[height=6cm]{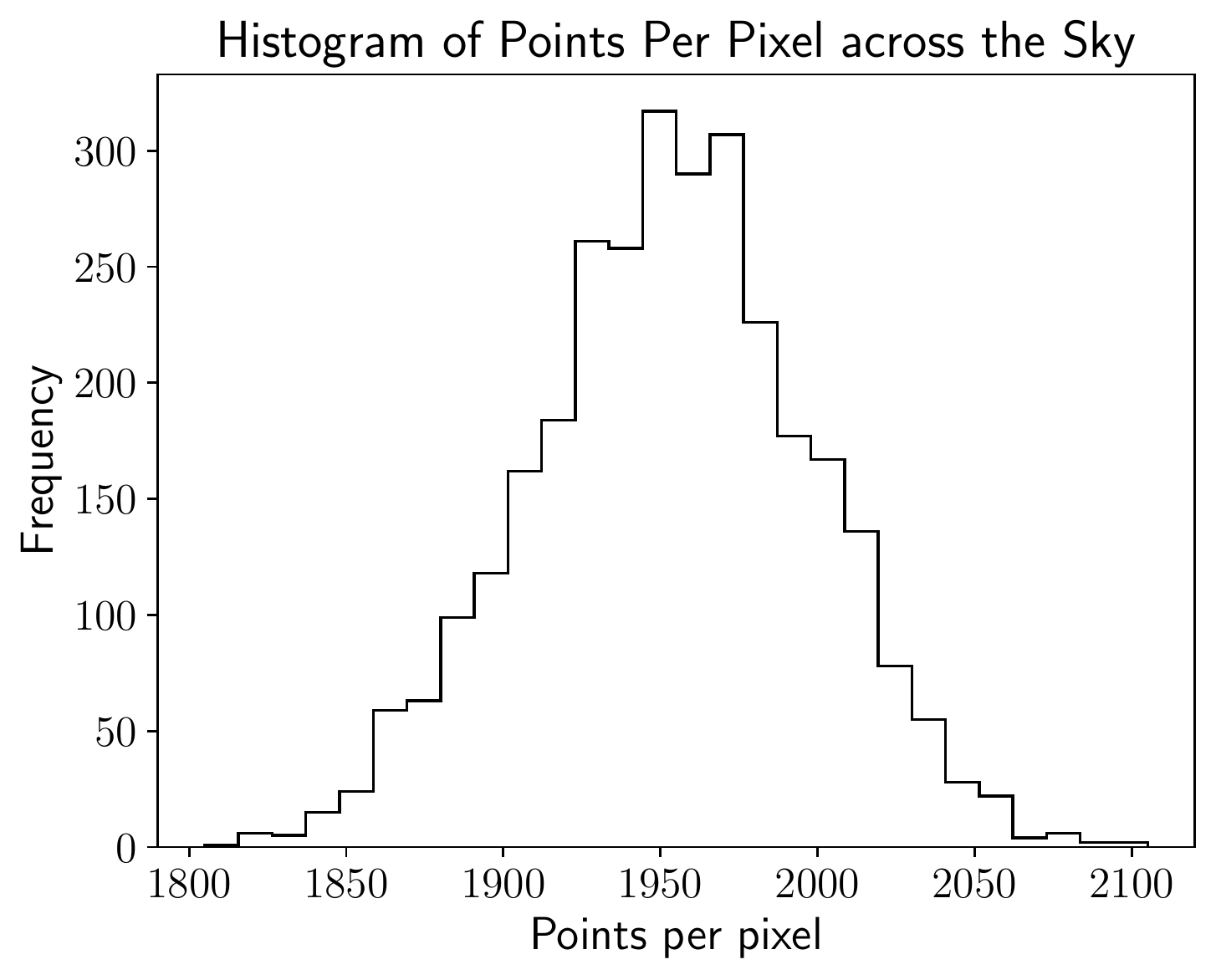}
    \caption{Binning of sources ($n = 6 \times 10^6$) into pixels for one run of the numerical simulation. \textit{Left:} Mollweide projection illustrating the density of points (sources) per pixel (mean density $\mu$ labelled), as displayed in galactic coordinates $(l,b)$. The galactic-plane mask of $|b| < 30^\circ$ is illustrated by the white region. \textit{Right:} Histogram of pixel number count densities for the density map shown to the left.}
    \label{fig:density-map}
\end{figure*}

\subsubsection{Principles of kinematic time dilation}
\label{sub:princ-kin-dil}
Separately to cosmological time dilation, our departure from the local Hubble flow will induce a directionally-dependent time
dilation component measurable in the characteristic time-scales of distant sources.
This effect is identical to the relativistic Doppler shift of sources arising from an observer's net motion towards or away from those sources.
Namely, this time dilation is described by
\begin{equation}
\Delta t_o = \gamma ( 1 - \beta \cos{\alpha} ) \Delta t_s
\label{eqn:doppler}
\end{equation}
where $\Delta t_s$ in the time-scale that would be observed if we were comoving observers,
$\Delta t_o$ is the observed time-scale given our peculiar motion $v$, $\beta = v/c $ and 
$\gamma = ( 1 - \beta^2 )^{-\frac{1}{2}}$ is the usual Lorentz factor. $\alpha$ is the angle subtended by the direction of the peculiar velocity and the source as measured in the observer's frame, and hence the resultant form of this 
kinematic time dilation is imprinted as a dipole over the sky.
Now, if the kinematic interpretation of the CMB is correct, then the velocity recovered from measurement of the `time dilation dipole' should correspond identically to the velocity inferred from the CMB temperature dipole.
This is the key principle behind our proposed test; like the matter dipole tests, it checks the assumption that the CMB defines a cosmic rest frame, and as such that our deviation from this frame matches with other observables, here being the time dilation of distant sources.

However, based on observations of the CMB, our kinematic departure from the local 
Hubble flow is of the scale $\beta\sim 10^{-3}$ and, by equation~\eqref{eqn:doppler}, corresponds to a maximal kinematic time 
dilation of $\approx \pm 0.1\%$; thus, the effect is subtle.
Moreover, as establishing the cosmic distance ladder was reliant on
the identification of standard candles and rulers, detecting the signature of kinematic 
time dilation will require the identification of standard clocks over the sky.
This catalogue would need to be sufficient in size to overcome the small scale of the effect, and, in the same vein, would presumably need substantial sky coverage.
Fortunately, upcoming large surveys of cosmically distant variable sources, which we describe in Section~\ref{Sec:observations}, offer the 
prospect of detecting kinematic time dilation against the backdrop of the isotropic cosmological signal.

In the context of our proposed test, these large catalogues of sources need robust determinations of intrinsic time-scales such that the subtle imprint of kinematic time dilation can be measured.
For instance, a proposed quasar catalogue will be bottlenecked by how readily time-scales can be assigned to the stochastic variability of quasar light curves.
As touched on earlier, the DRW model parametrizes these aperiodic fluctuations with two variables: the characteristic time-scale $\tau$ and variability amplitude $\sigma$, in which case the covariance matrix of the signal can be described by $S_{ij} = \sigma^2 \exp ( - | t_i - t_j | / \tau)$ for the $i$-th and $j$-th timesteps $t_i$ and $t_j$ \citep{macleod2010}.
In this way, the time-scale, which is in principle affected by kinematic time dilation, can be extracted from a light curve.
This is not an exhaustive way to describe quasar variability; more generally, the DRW is the simplest model belonging to the family of Continuous Autoregressive Moving Average Models (CARMAs), which are used to provide statistical, not physical, descriptions of stochastic processes \citep{Kelly_2014,sheng2022}.
However, there is evidence to suggest DRWs in certain circumstances are insufficiently complex to account for quasar variability, and a quasar light curve may be better described by the higher-order Damped Harmonic Oscillator \citep[DHO;][]{kasliwal2015,kasliwal2017}.
The point is that there are statistical models which exist in describing these fluctuations, albeit with varying efficacy.
We may thus associate a light curve with a characteristic time-scale, and in this manner search for the imprint of kinematic dilation across large catalogues.

\subsection{Mock test procedure}
\subsubsection{Simulated catalogue}
How would we extract the signal of Earth's motion from a catalogue of time-scales?
To showcase this, we generated a simulated sample of time-scales ascribed to hypothetical sources, which are assumed to all be identical.
Though this example is straightforward and will not be entirely reflective of observational reality, it is none the less illustrative of the process required to measure the time dilation dipole.
Now, each time-scale $\tau_0$ in this simulation is attributed with Gaussian noise, all drawn from a distribution $\mathcal{G}$ with common width $\Delta \tau_0$ such that the $i$-th timescale is given by
\begin{equation}
    \tau_i = \tau_0 + \mathcal{G}_i(\mu = 0, \sigma = \Delta \tau_0). \label{eqn:noise}
\end{equation}
The distribution of sources in the cosmic rest frame is uniform across the sky, as per the cosmological principle.
The observer's peculiar motion, however, will imprint a dipole on the simulated catalogue: there will not only be an enhancement in number count density along the line of motion (relativistic aberration), but also a dilation in time-scale, such that sources along the line of motion are associated with a shorter time-scale.
This is again as described by equation~\eqref{eqn:doppler}.
In the observer's coordinates, each source is therefore characterised by different angles $\alpha$ between the line of motion and source, as well as different time-scales $\tau'$.

We then binned the sources over the sky using the Python package \textsc{healpy}, an implementation of the HEALPix\footnote{\url{https://healpix.sourceforge.io/}} scheme \citep{Gorski2005,Zonca2019}.
With $N_\text{side} = 16$, $\approx 3000$ pixels ($n_{\text{pix.}}$) of equal area are generated, as illustrated for one particular run in the left pane of Fig.~\ref{fig:density-map}, there leading to a distribution of pixel densities shown in the right pane of the same figure.
For each pixel $\mathcal{P}_j$, we then compute an average $\bar{\tau}'_j$ and standard deviation $\sigma_j$ over all time-scales within that pixel, and apply a conservative galactic plane mask, discarding pixels within $|b| < 30^\circ$ \citep[see e.g.][]{2021ApJ...908L..51S}.
Accordingly, in this simulated catalogue, we are looking for how the average time-scale ascribed to each pixel changes as a function of location on the sky.
The key deductive step is then to compare the velocity inferred from this sample with that determined from the CMB dipole.

\begin{figure}
    \centering
    \includegraphics[width=0.48\textwidth]{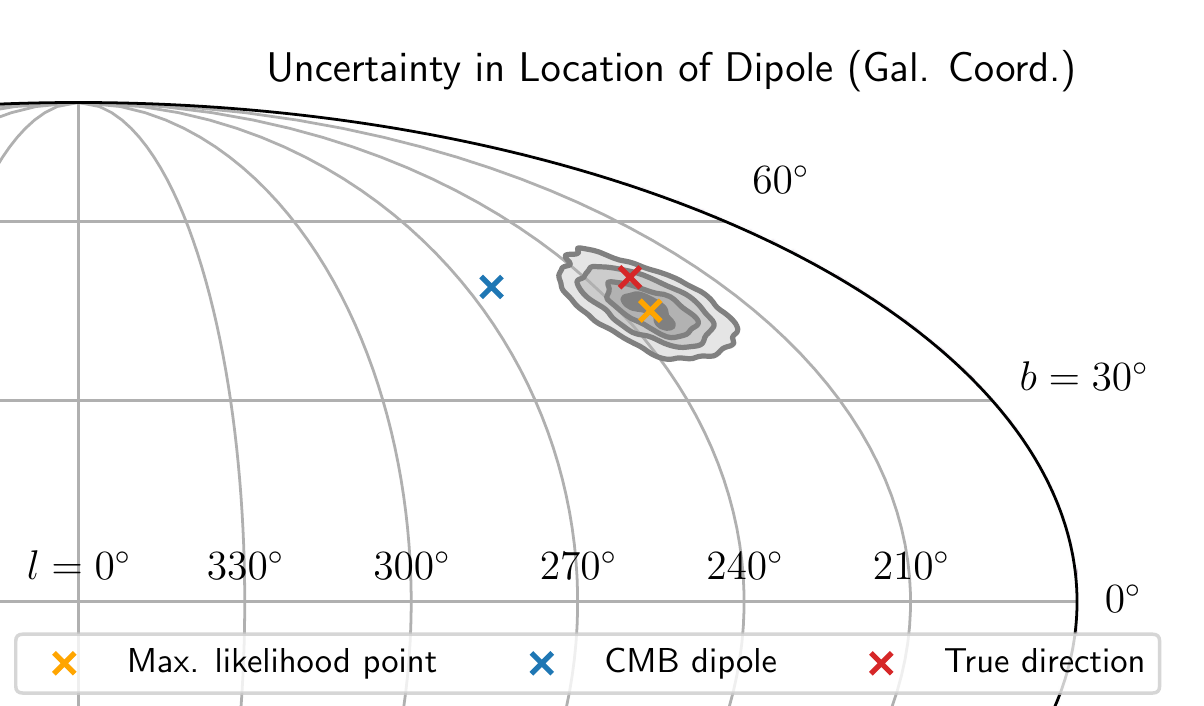}
    \caption{Uncertainty in the direction of the time dilation dipole from a random run of the sample ($n = 1.6 \times 10^7$ sources). The 2D marginal posterior distribution for the direction of the dipole is projected (Mollweide) onto the sky with galactic coordinates. The contours represent intervals of posterior density (10\%, 40\%, 65\%, 85\%), with the direction of maximum likelihood given by the yellow cross. The blue and red crosses show the direction of the CMB dipole and the simulated true direction of motion respectively.}
    \label{fig:sky-uncertainty}
\end{figure}

Thus, at the initial source generation, we imprint kinematic time dilation on the time-scales as described by motion $30^\circ$ off-alignment with the CMB dipole but approximately the same magnitude ($v = 0.001c$).
We do so to understand the sensitivity of our proposed test; if the time dilation dipole is in fact not aligned with the CMB dipole, how many sources would be necessary to reach this conclusion?
Put differently, we are verifying whether or not this time dilation signal can be extracted from the sample, which necessarily is a question of whether or not it can be sufficiently resolved from the CMB signal.
As clarification, Fig.~\ref{fig:sky-uncertainty} illustrates this imprinted `true direction' in the vicinity of the direction of the CMB dipole and the most likely direction of the time dilation dipole, as determined from one random instance of the sample.

As a separate experiment, we turned to the scenario of alignment between the time dilation dipole and the CMB dipole.
If this is the case, then the only distinguishing feature, as far as our test is concerned, is the magnitude of either observable.
Accordingly, here we altered the time-scales of each source according to motion aligned with the CMB dipole but with a velocity in excess.
The question then becomes: to what degree does this change the number of sources required before the time dilation dipole can be resolved?

\subsubsection{Bayesian analysis and hypothesis testing}
\begin{figure*}
    \centering
    \includegraphics[width=0.8\textwidth]{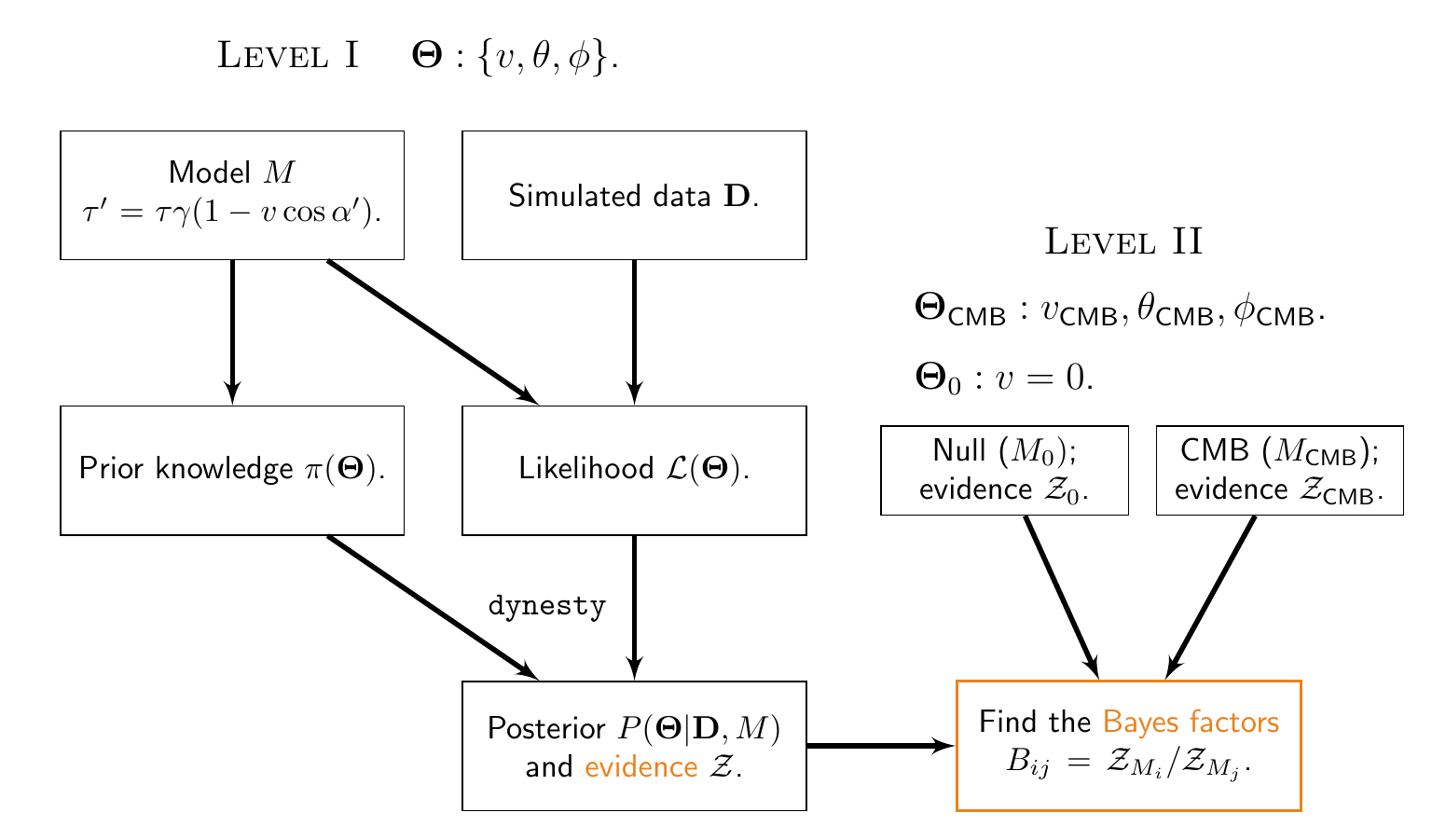}
    \caption{Flowchart summarising the key elements of our proposed test with Bayesian methodology. On the left hand side, at the first level of Bayesian inference (Level I), the best-fitting parameters for the observer's motion from the simulated data $\mathbf{D}$ are determined. The evidence $\mathcal{Z}$ is then computed. These are all determined through \textsc{dynesty}'s implementation of the NS algorithm. At the second level of Bayesian inference (Level II), the evidence for each hypothesis is compared through finding Bayes factors, as given in equation~\eqref{eqn:bayes-factors}.\label{fig:bayesian-flowchart}}
\end{figure*}
\noindent To carry out these statistical evaluations, we employ a Bayesian approach, and so need to populate the terms of Bayes's theorem. In this case, the posterior probability distribution is given by
\begin{equation}
    P( \mathbf{\Theta} | \mathbf{D}, M ) = \frac{\mathcal{L}(\mathbf{D} | \mathbf{\Theta}, M) \pi(\mathbf{\Theta} | M)}{\mathcal{Z}(\mathbf{D} | M)}
\end{equation}
for set of parameters $\mathbf{\Theta}$ pertaining to model M and observed data $\mathbf{D}$.
Here, $\mathcal{L}$ is the likelihood function, $\pi$ is the prior function and $\mathcal{Z}$ is the marginal likelihood or Bayesian evidence.
In our approach, the set of parameters interrogated are $\mathbf{\Theta} = \{ v, \theta, \phi \}$, representing the observer's peculiar velocity, as well as the polar and azimuthal components of the direction of that peculiar velocity in an arbitrary spherical coordinate system.

The likelihood of obtaining the time-scales $\bar{\tau}'_1,\bar{\tau}'_2,\ldots,\bar{\tau}'_{n_{\text{pix.}}}$ for each pixel is given by
\begin{equation}
    \mathcal{L} = \prod_{j=1}^{n_{\text{pix.}}} P(\bar{\tau}'_j | \mathbf{\Theta} ) =\prod_{j=1}^{n_{\text{pix.}}}\mathcal{G}_j (\bar{\tau}'_j | \, \mu = \tau'_j, \sigma = \sigma_j).
\end{equation}
That is to say, the total likelihood is the product of the likelihood for all pixels, each of which is obtained by computing the value of a Gaussian at point $\bar{\tau}'_j$ where the Gaussian is centred at $\tau'_j$ and has width $\sigma_j$.
Here, as mentioned earlier, we assume \textit{a priori} knowledge of the true rest-frame time-scale $\tau'_j$ for each pixel.
For the prior function, a uniform distribution for $\theta$ and $\phi$ across the entire sphere was adopted, as well as a uniform distribution for $v$ between 0 and $0.01c$.

\begin{figure*}
    \centering
    \includegraphics[width=0.48\textwidth]{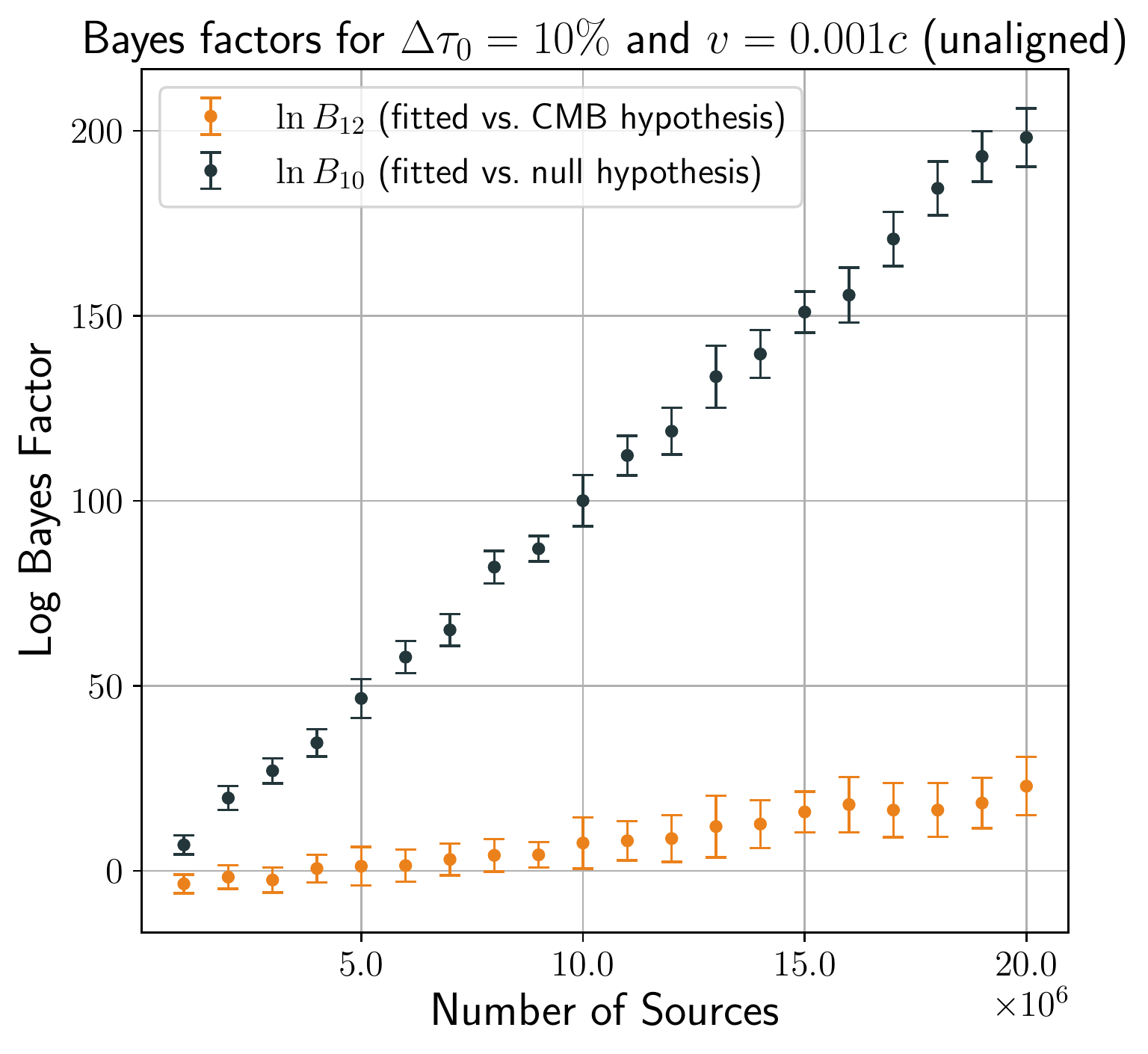}
    \hfill
    \includegraphics[width=0.48\textwidth]{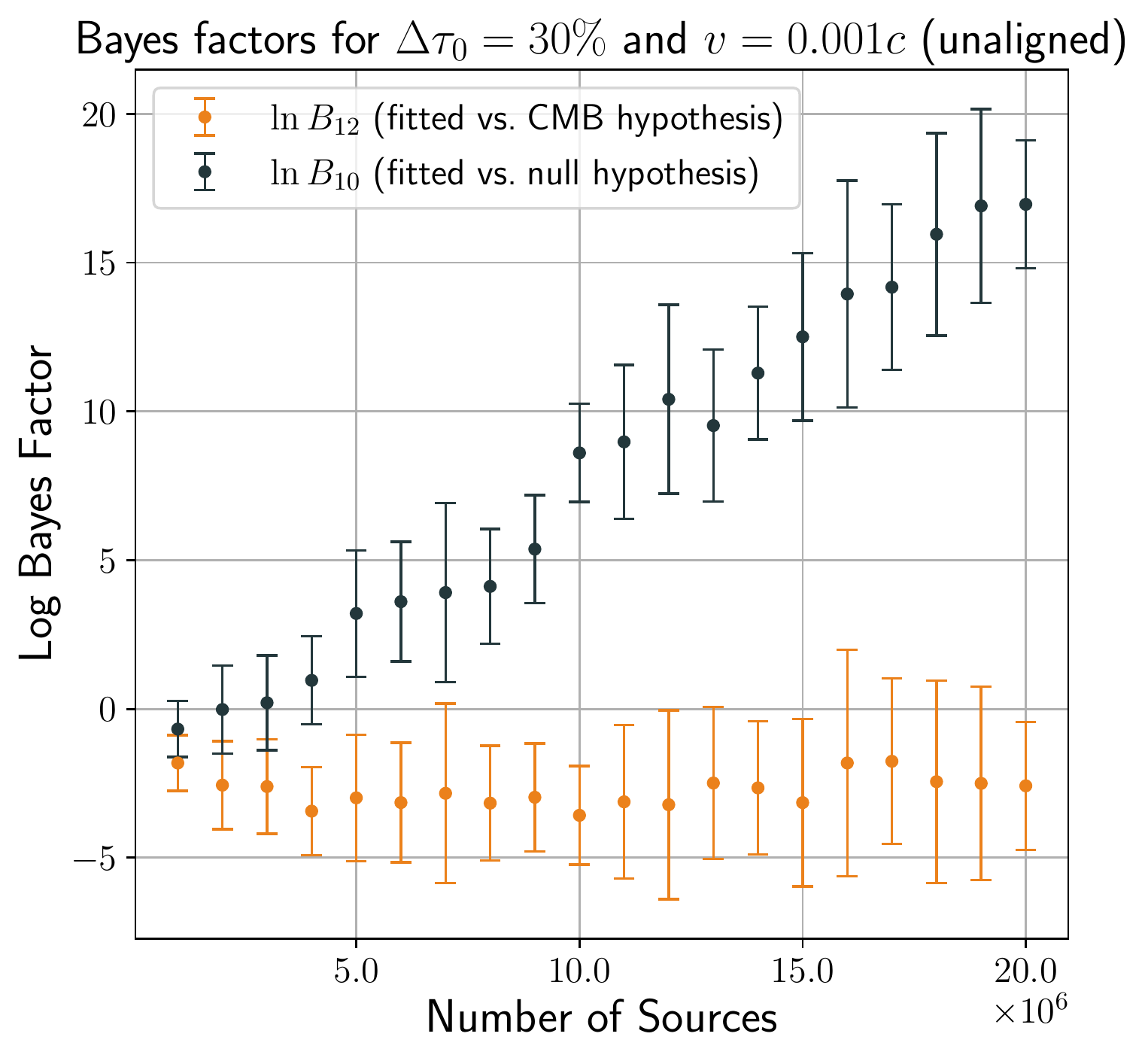}
    \includegraphics[width=0.48\textwidth]{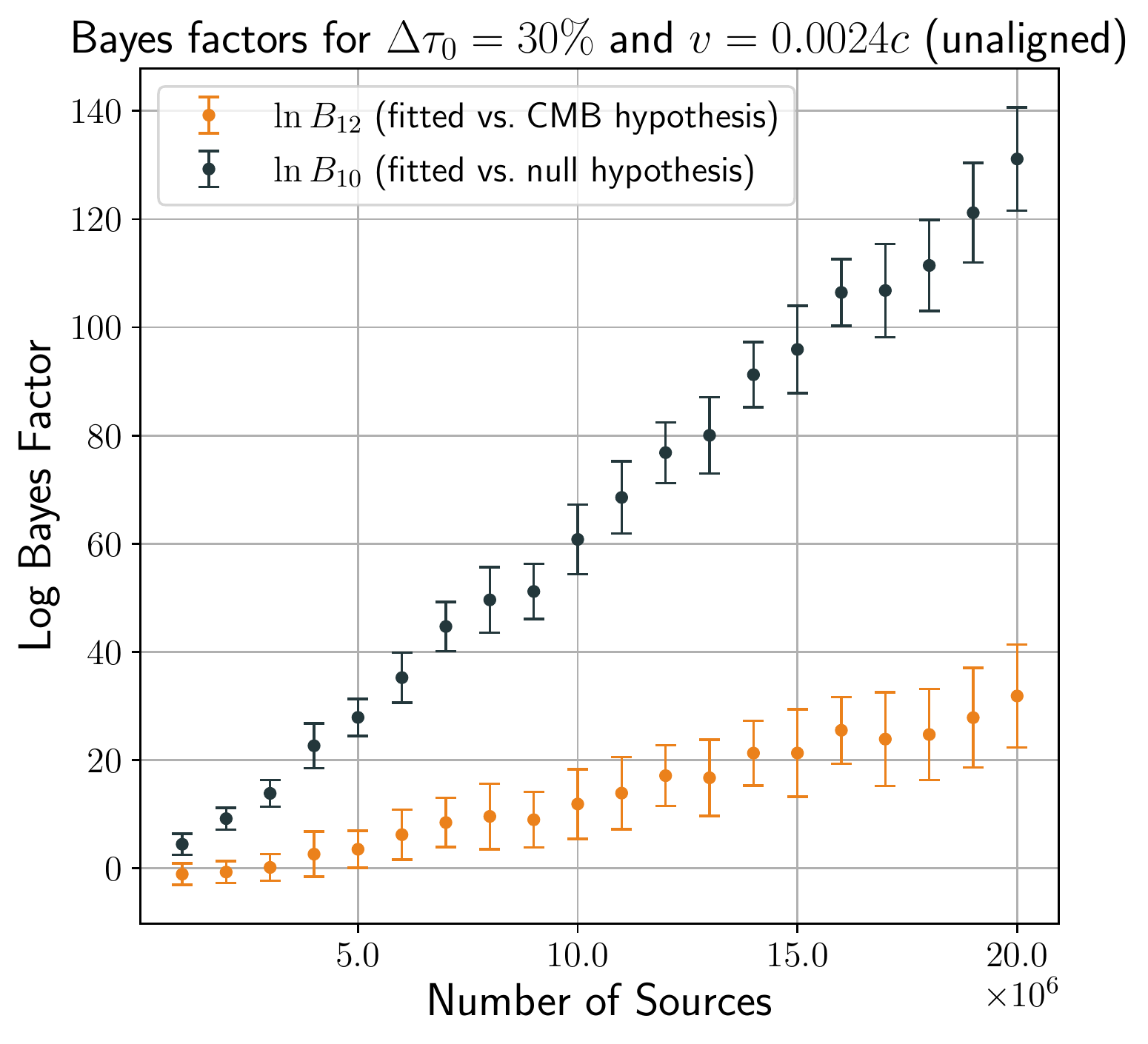}
    \hfill
    \includegraphics[width=0.48\textwidth]{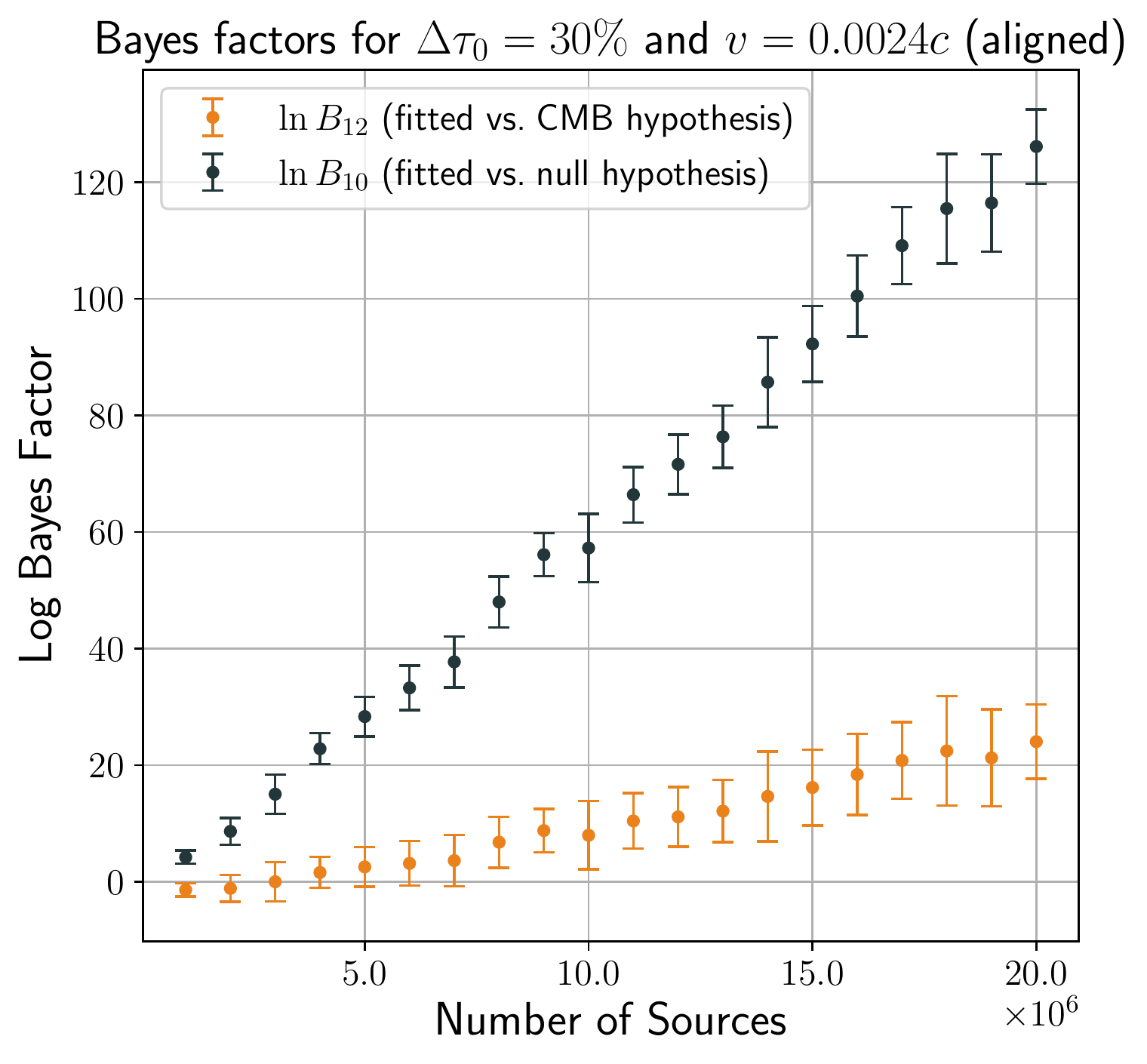}
    \caption{Bayes factors against the number of sources in the simulated catalogue. Here, the observer's peculiar velocity, the width of the Gaussian noise added to each source (see equation~\eqref{eqn:noise}) and the true direction of the time dilation dipole is varied. \textit{Top left:} $\Delta \tau_0 = 10\%$, $v=0.001c$, $30^\circ$ off-alignment with the CMB dipole. \textit{Top right:} $\Delta \tau_0 = 30\%$, $v=0.001c$, $30^\circ$ off-alignment with the CMB dipole. \textit{Bottom left:} $\Delta \tau_0 = 30\%$, $v=0.0024c$, $30^\circ$ off-alignment with the CMB dipole. \textit{Bottom right:} $\Delta \tau_0 = 30\%$, $v=0.0024c$, aligned with the CMB dipole.\label{fig:bayes-factors}}
\end{figure*}

This leaves us with the marginal likelihood or evidence, which is traditionally challenging to compute as it represents an integral over all parameter space $\Omega_{\mathbf{\Theta}}$: $\mathcal{Z} = \int_{\Omega_{\mathbf{\Theta}}} \mathcal{L} (\mathbf{\Theta}) \pi (\mathbf{\Theta}) \, d \mathbf{\Theta}$.
In this work, we deployed the package \textsc{dynesty} \citep{dynesty-v1.2.3}, a Python implementation of the Nested Sampling algorithm \citep[NS;][]{skilling2004,skilling2006} which evaluates the evidence integral and returns the posterior probability distribution.
This gives us a natural language to evaluate competing models or hypotheses.
In particular, models which better explain the data will generally have a larger evidence as the likelihood function is optimised.
However, complexity will be penalised; models with excessive parameters will waste parameter space, reducing the value of $\mathcal{Z}$.
These competing interests are encapsulated in the evidence, and we can therefore compare the support for different hypotheses.
In this study, these are the following:
\begin{itemize}
    \item The null hypothesis $M_0$, in which the observer has no peculiar velocity and as such there is no time dilation dipole imprinted on the sample;
    \item The fitted hypothesis $M_1$, in which we determine (optimise) the magnitude and direction of the observer's peculiar velocity given the sample; and,
    \item The CMB hypothesis $M_2$, in which the magnitude and direction of the observer's peculiar velocity is supposed to coincide with that as determined from the CMB dipole.
\end{itemize}
%
%
%
We compare the support for each hypothesis by evaluating Bayes factors. These are the ratios of the model evidences, and they are traditionally used to understand which model is preferred given a set of data, weighing both models' explanatory power and complexity. Explicitly, in our case they are:
%
\begin{equation}
    B_{12} = \frac{\mathcal{Z}_1}{\mathcal{Z}_2} \quad \text{and} \quad B_{10} = \frac{\mathcal{Z}_1}{\mathcal{Z}_0}. \label{eqn:bayes-factors}
\end{equation}
This will give an indication of how much the fitted hypothesis is favoured over the CMB and null hypotheses respectively, which can be interpreted qualitatively using the values quoted in \citet{kass1995}.
For example, a value of $B_{12}$ fractionally greater than 1 means that the fitted hypothesis is only just favoured over the CMB hypothesis.
The intuition for our test is that the Bayes factors will be strongly correlated to the number of original sources $n$; as $n$ increases, a time dilation dipole is more robustly extracted from the data.
We summarise the foregoing in the flowchart of Fig.~\ref{fig:bayesian-flowchart}, since, at least in a Bayesian setting, these are basic steps required to extract the time dilation dipole and perform statistical inference.

\subsection{Findings}
They key results are shown in Fig.~\ref{fig:bayes-factors}.
Here, the natural logarithm of the Bayes factors in equation~\eqref{eqn:bayes-factors} are plotted as a function of the number of points or sources $n$ in the simulated catalogue before the mask is applied.
The error bars represent $1 \sigma$ of the logarithmic Bayes factors after repeated trials, in general being $\approx 20$ for each point.
We note that, as expected, the Bayes factors are strongly correlated with $n$; simply, a larger sample of time-scales equates to a clearer resolution of the time dilation dipole over the CMB dipole.
This has a clear exception for $\ln B_{12}$ where $\Delta \tau_0 = 30\%$ and $v=0.001c$ (top right pane), which we will discuss below.
We also note that larger noise in the original sample of time-scales implies a larger variance of the Bayes factors for the same speed.
We again leave the qualitative interpretation of the Bayes factors themselves to \citet{kass1995}, stressing that these values reflect the quantification of an abstract continuous variable into discrete normative categories, so are not totally prescriptive.
However, they do suggest the point at which noteworthy support emerges for a model, namely past $\ln B_{ij} = 1$.

%

With all this in mind, we turn now to the case of the true direction of motion being unaligned with the CMB dipole by $30^\circ$. These are represented by the top left, top right, and bottom left panels of Fig.~\ref{fig:bayes-factors}.
We infer that the fitted hypothesis has more explanatory power than the other hypotheses where $n \gtrsim 10^7$ for $\Delta \tau_0 = 10\%$ (the top left pane of Fig.~\ref{fig:bayes-factors}).
After masking, this corresponds to $\approx$ 5 million sources.
Moreover, the fitted hypothesis always has more support than the null hypothesis for any $n$.
However, where $\Delta \tau_0 = 30\%$ (the top right pane of Fig.~\ref{fig:bayes-factors}), although after a certain $n$ the fitted is favoured over the null, the fitted is \textit{never} preferred over the CMB hypothesis, at least for the sample size range tested.
Neither does increasing $n$ imply stronger (larger) evidence for the fitted hypothesis.
We turn to the implications of this below.

In line with the tension within the literature, we also investigated the case of the observer's peculiar motion being twice as large ($v = 0.0024c$) than that as determined from the CMB dipole, with the same direction of motion as above.
The results are summarised in the bottom left pane of Fig.~\ref{fig:bayes-factors}.
There, we note that an $\approx$ factor of 2 increase in the observer's peculiar speed confers the ability to resolve the time dilation dipole at $\Delta \tau_0 = 30 \%$.
Support emerges for the fitted over the CMB hypothesis where $n \gtrsim 5 \times 10^6$, corresponding to about 2.5 million sources after masking.
This is a substantial reduction from the value quoted earlier at the lower speed and uncertainty.
Accordingly, if the suggestion of an excessive dipole magnitude reflects physical reality, the effect of kinematic time dilation would be more readily discerned.

What if the direction of motion is aligned with the CMB dipole, but has a velocity in excess?
In this case, the results are illustrated in the bottom right pane of Fig~\ref{fig:bayes-factors}, noting that $\Delta \tau_0 = 30\%$ and $v = 0.0024c$.
This enables quick comparison with the bottom left pane, which differs only in the direction of motion.
We observe that it takes more sources -- specifically $n \gtrsim 8 \times 10^6$ (4 million after masking) -- before the fitted model is favoured over the CMB hypothesis.
Accordingly, we deduce that more sources are needed to reject the CMB hypothesis if the observer's motion is in fact aligned with the CMB dipole, but has a larger than anticipated velocity.

The foregoing gives an order of magnitude estimate for the number of sources with characteristic time-scales needed to perform our proposed test.
Evidently, the key limiting factor is $\Delta \tau_0$.
What is the physical interpretation of this value?
In practice, with a real catalogue of sources, we would divide the sky into cells, similarly to as we have hitherto done, but bin over a redshift range to compensate for cosmological time dilation (see Section~\ref{sub:cosm-time-dil}).
Each bin contains a distribution of time-scales, which, assuming they are sufficiently large, we expect to match with other bins across the sky at the same redshift range \textit{but for the effect of kinematic time dilation}.
In other words, we could extract a median or mean time-scale from this characteristic distribution and check this against other locations on the sky, in principle discerning the imprint of kinematic time dilation by how this location parameter is impacted.
$\Delta \tau_0$ is analogous to the uncertainty in fixing a value to this location parameter; the greater it is, the less confidently we can measure the effect of kinematic time dilation.
Thus, our test is highly sensitive to how precisely an observed distribution of time-scales can be characterised.
Specifically, this is evinced by the model's inability to resolve the imprint of time dilation at $\Delta \tau_0 = 30\%$ and $v = 0.001c$ (top right pane of Fig.~\ref{fig:bayes-factors}).

\section{Observational Prospects}
\label{Sec:observations}

Catalogues of quasars are a natural choice for our proposed test. Yet, we mentioned in Section~\ref{sub:cosm-time-dil} that, as it stands presently, AGN studies may not involve sufficient sampling to reveal the effect of cosmological time dilation on characteristic time-scales, let alone kinematic time dilation. 
There is already a body of work which indicates, at least for DRW models, that a light curve needs to be sampled for a duration at least ten times the rest frame time-scale to enable its accurate determination \citep[see e.g.][]{kozlowski2017,kozlowski2021,sheng2022}.
Keeping this accuracy in mind, there is the additional suggestion in \citet{kozlowski2017} that, assuming the duration of sampling is sufficient, the cadence needs to be at least shorter than the time-scale.

For instance, the analysis in \citet{stone2022} was conducted with a small ($\approx$ 200) sample of quasars measured over $\approx$ 20 years, achieved by combining data from the Sloan Digital Sky Survey (SDSS), Pan-STARRS1 \citep{chambers2016}, the Dark Energy Survey \citet{hartley2022} and a follow-up study with the Dark Energy Camera on the Cerro Tololo Inter-American Observatory 4m telescope.
The authors recovered DRW parameters by modelling the light curves with Gaussian processes and noted that, even with a baseline of this duration, there was still some correlation between the intrinsic DRW timescale and the baseline length.
This suggests that the quasars may still be insufficiently sampled to constrain the timescale (i.e. the timescale is more than 20\% of the baseline duration), or that the DRW model is too simple to properly account for the stochastic variability of the sampled quasars.

Available measurements therefore leave much to be desired in terms of constraining DRW parameters, despite the significant volume of objects in surveys like SDSS.
This presents an issue for our approach, since we need to first accurately characterise a source's variability before attempting to discern the effects of time dilation.
However, the upcoming Legacy Survey of Space and Time (LSST) at the Vera Rubin Observatory represents the next leap in studies of AGN variability.
As the `most ambitious survey currently planned in the optical', the Wide-Fast-Deep (WFD) survey will monitor a catalogue of quasars larger than 10 million over a period of ten years \citep{LSST2019}.
As the authors of \citet{stone2022} note, seamless integration of the LSST catalogue and available data will provide an excellent sample to study the applicability of the DRW model or, for example, high-order CARMA models.
In light of the order-of-magnitude estimates we quoted in the previous section for the number of sources/timescales, this means that while our novel test is likely not feasible with current data, the future is bright.
In particular, we imagine that LSST's completion will yield a candidate catalogue for our proposed probe of the cosmological principle.

Still, there are additional challenges to overcome, for example the cadence strategy deployed by WFD.
LSST is a substantial project with a diverse spectrum of scientific goals; accordingly, a host of different observing strategies have been proposed.
Thus the optimisation of this strategy is a `complex challenge' \citep{lochner2022}.
As it currently stands, LSST will have a main survey footprint of 18,000\,deg$^2$ with a median of 825 30-second visits per field in all filters \citep{jones2021}.
This leaves substantial flexibility for how the details of the key survey parameters, like cadence, are optimised. These have not as of yet been finalised.
A select few of these different approaches are studied in \citet{sheng2022}.
There, the authors used a novel machine learning approach to recover DRW and DHO parameters from simulated quasar light curves, sampled with some of the proposed observational strategies for LSST.
They cautioned that the long seasonal window, as part of all proposed cadences, is the key limiting factor if the priority is for the accuracy of parameter extraction.
Whilst LSST's Deep Drilling Fields (DDF) will have a higher and more uniform cadence and so may solve this issue, our test prioritises a larger volume of objects, and so the smaller samples of the DDFs will probably not be suitable.
Thus, WFD's catalogue is the most promising for our test.
It is also conceivable that in the time-span between now and the completion of LSST's ten-year program, substantial headway will be made by the AGN community in modelling light curves.
This may resolve some of the issues mentioned above, leading to an optimal approach in DRW timescale extraction, or more sophisticated models which better explain quasar variability.
In essence, this leads us to the view that the time dilation dipole will be detectable within the vast bulk of sources monitored by LSST.


\section{Conclusions}
\label{Sec:conclusions}
Observations of the CMB clearly reveal a dipole in temperature, interpreted 
as kinematic Doppler shift due to our departure from the local Hubble flow. However, there is
some discord in the magnitude and direction of our local motion drawn from other observational probes, leading to a questioning of the validity of cosmological principle.
If there is lingering doubt on the assumption of an isotropic and homogeneous universe, then this is a direct challenge to the prevailing paradigms of modern cosmology.
In this sense the tension is a serious issue that cannot be overlooked.

This paper has presented a new potential probe of our departure from the local Hubble flow through the detection 
of the kinematic time dilation of cosmologically distant sources.
In preparing a numerical simulation formulated using these physical principles, we gave an order of magnitude estimate for the number of sources required in a typical catalogue to detect the time dilation dipole.
The critical point is that this illustrates the feasibility of our proposed test, and also provides a framework for how the test may be approached in the future.
This is especially relevant in light of current observational prospects.
LSST will reshape AGN studies, creating a catalogue with unprecedented size and density of observations.
This will improve our ability to characterise the stochastic variability of quasar light curves with intrinsic time-scales, which are expected to be affected by kinematic time dilation.
In this manner, we may add to the suite of tools used to interrogate the core assumptions of modern cosmology, either showing them to be sound or lending further evidential support for the growing anomaly.
In either case, we hold the fundamental assumptions we have of the universe up to scrutiny, lighting the way for the cosmological models of the future.


\section*{Acknowledgements}
We thank the anonymous referee for their insightful comments. This work made use of the Python packages \textsc{dynesty} \citep{skilling2004, skilling2006, dynesty-v1.2.3}, \textsc{healpy} \citep{Gorski2005,Zonca2019}, \textsc{numpy} \citep{harris2020}, \textsc{matplotlib} \citep{hunter2007}, \textsc{pandas} \citep{mckinney2010,reback2020} and \textsc{scipy} \citep{scipy2020}. 

\section*{Data Availability}
The data used in this study will be made available with a reasonable request to the authors.



\bibliographystyle{mnras}
\bibliography{paper} 






\bsp	
\label{lastpage}
\end{document}